# *Chronos*

Take the pulse of our Galactic neighbourhood
**After *Gaia*: time domain information, masses and ages for stars**

A white paper for the ESA Voyage 2050 long-term plan

**Contact scientist:** Eric Michel, Observatoire de Paris, LESIA, Université PSL
Eric.Michel@obspm.fr

**Core proposing team :**

- **Eric Michel, Kevin Belkacem, Benoit Mosser, Reza Samadi** – LESIA, Observatoire de Paris, Université PSL, CNRS, Sorbonne Université, Université de Paris, Paris France
- **Misha Haywood, David Katz** – GEPI, Observatoire de Paris, Université PSL, CNRS, Paris, France
- **Benoit Famaey** – Université de Strasbourg, CNRS, UMR 7550, Observatoire astronomique de Strasbourg, 11 rue de l'Université, 67000, Strasbourg, France
- **Tiago Campante, Mario J.P.F.G. Monteiro, Margarida Cunha** – Instituto de Astrofísica e Ciências do Espaço, Universidade do Porto, Rua das Estrelas, 4150-762 Porto, Portugal — Departamento de Física e Astronomia, Faculdade de Ciências da Universidade do Porto, Rua do Campo Alegre, s/n, 4169-007 Porto, Portugal
- **Andréa Miglio** – University of Birmingham, Birmingham, United Kingdom
- **Rafael Garcia** – Département d'Astrophysique, IRFU/DRF/CEA Saclay, L'Orme des Merisiers, bat. 709, 91191 Gif-sur-Yvette Cedex, France
- **Hans Kjeldsen** – Stellar Astrophysics Centre. Department of Physics and Astronomy. Aarhus University, Denmark
- **Juan Carlos Suarez** – Ramon y Cajal fellow at Física Teórica y del Cosmos Dept. University of Granada Facultad de Ciencias. 18071. Granada. Spain
- **Sébastien Deheuvels, Jerôme Ballot** – IRAP, Université de Toulouse, CNRS, CNES, UPS, Toulouse, France

# *Chronos*

**Take the pulse of our Galactic neighbourhood**
After *Gaia*: time domain information, masses and ages for stars

## 1) Introduction:

### Understanding our Galaxy with *Gaia* and time domain constraints.

Galactic archaeology has entered a new era with the data provided by the *Gaia* mission, allowing to get new insights into fundamental questions about the build-up of our Galaxy and its different stellar components. For instance, is the actual backbone of our Galaxy a "disky" stellar component, or was there an early in-situ stellar halo? What is the role played by early mergers and accretions of satellite galaxies? What was (and still currently is) the respective role of external accretions and internal dynamical instabilities (bar, spiral arms), as well as their interplays, in shaping the star formation history and the dynamics of the different stellar components? Answering all these questions requires determining the ages of stars.

The *Gaia* mission (*Gaia* Collab. et al. 2018a) is providing us with an unprecedented insight of our Galaxy. Completed by the associated large spectroscopic surveys (such as APOGEE and WEAVE), *Gaia* is characterizing in detail and with a unique precision, positions (including distances), movements, effective temperatures, luminosities, chemical compositions for about a billion stars over a large portion of our Galaxy and nearby satellite galaxies (e.g., *Gaia* Collab et al. 2018b).

These precise data shed a new light on the above fundamental questions about how our Galaxy has formed and evolved. They reveal unexpected trends, suggesting for instance that a large fraction of the oldest stars in the Galaxy have "disky" orbits (Sestito et al. 2019), that the stellar halo is almost fully accreted (e.g., Di Matteo et al. 2018) and that these accreted components in turn played an important role in shaping the in-situ "disky" components (e.g., Helmi et al. 2018), while it appears that more recent accretions are still influencing disk dynamics today (Antoja et al. 2018, Laporte et al. 2019). The respective roles of such external perturbations and internal instabilities such as the Galactic bar, the structure of which also being constrained with *Gaia* (e.g., Monari et al. 2019) , and the interplay between these external and internal dynamical mechanisms, are however still under investigation. To reach a more precise view of the situation, we however absolutely need to enrich the very precise astrometric-kinematic-chemical view of the Galaxy provided by *Gaia* and large spectroscopic surveys with high precision ages which can be provided through seismic information (or more generally time domain information).

A large fraction of stars shows multi-periodic photometric variations due to stellar pulsations, but also rotational modulation, granulation... These phenomena and the measurement of their photometric signature carries valuable information of different nature than the astrometric-kinetic-chemical one, like mass, age and (unprojected) rotation estimates. With CoRoT (Baglin et al. 2006) and *Kepler* (Borucki et al. 2010), we have learned to derive this information from photometric light curves of tens of thousands of stars and we can foresee an upgrade of this capacity in the number of stars but also in type of stars and in type of applications.

**This is why we are firmly convinced that it will be a top priority for the scientific community in 2035-50 to improve our view and understanding of our Galaxy, by adding a time domain dimension to the *Gaia* Survey. *Chronos* will complete the *Gaia* survey with high-temporal resolution for a million stars with 8≤$m_V$≤11.**

-in Sect.2 we discuss the nature of the information brought by the variability dimension and show why it is original and complementary to information on stars brought by *Gaia* and spectroscopic surveys and why a strong potential synergy exists. We also stress the specific constraints characterizing time domain measurements with a special emphasis on the project considered here.

-in Sect.3 we show how our understanding of the Galaxy will progress with the variability information. In order to be practical, we realise this study in the framework of a specific example of survey focussing on red giants for which an instrumental option is proposed in the next Section.

-in Sect.4 we describe a possible option for an instrument and a mission profile capable of providing such a variability survey. We compare this instrument with existing or planed missions and show how unique and well suited it is for our purpose.

- Section 5 addresses possible improvements for the scientific objective and technical challenges for the mission concept presented here.

-in Sect.6 we summarize and conclude in a broad scientific perspective.

## 2) Specific nature of time domain characterization and synergy with *Gaia* and large spectroscopic surveys

As already mentioned, photometric variability exists at different amplitude levels and different time scales in a large proportion of stars. This variability can be due to stellar pulsations, but also to other phenomena like rotation modulation associated with activity or binary tide effects, granulation…

### 2a) Specificity of time domain information

**Considering stellar pulsation and seismic information,** the case of red giant stars is very illustrative of the power of this approach. From the observation of several tens of thousands of red giants by CoRoT and *Kepler*, we have learned to understand their oscillation spectra and to interpret them in terms of a few parameters, the so-called seismic indices (e.g. Mosser & Appourchaux 2009, Kallinger et al. 2010, Bedding et al. 2011, Mosser et al. 2012):

– **Δν:** named the large separation, characterizes the first-order regularity in frequency of acoustic oscillation modes. This seismic index is tightly linked with the stellar mean density (Eddington 1919). It is a very valuable piece of information about the individual stars, because, via the strong relation between radius and mass it imposes, it brings the possibility to characterize the stars in terms of mass, which is not measured with *Gaia* and only very poorly (via log g) in the spectroscopic analysis.

– **ν$_{max}$:** the frequency of maximum pulsation amplitude is strongly linked with the surface gravity, and, to a lesser extent with the effective temperature (Belkacem et al. 2011). It is another very valuable piece of information about the individual stars, because it gives accurate log*g* with a precision much better than spectroscopic analysis. When spectroscopic estimates of log*g* are known within a factor 2 or 3 for field stars, so-called 'seismic log g' are measured with a precision better than 2% (Bruntt et al. 2012)

The seismic indices Δν and ν$_{max}$, combined with the spectroscopic measurements of effective temperature, are used to provide stellar masses and radii based on simple seismic scaling relations. They have been used to derive precise estimates of individual radii and masses for about 55k red giant stars observed with CoRoT (~10k), *Kepler* (~20k), K2 (~10k), and TESS (~15k) (e.g. De Assis Peralta et al. 2018, Pande et al. 2018, Jie at al. 2018). These seismic indices are available online (see e.g. the Stellar Seismic Indices data base http://ssi.lesia.obspm.fr/). They are also regularly used in an iterative spectroscopic analysis to produce more precise gravity estimates allowing more precise chemical composition analysis.

For several thousands of these red giants, with long enough observations, it has been possible to measure a third seismic index (Vrard et al. 2016):

– **ΔP** is the period spacing for g-dominated oscillation dipole modes. It asymptotic value is measuring the size of the inner radiative core of red giants (Montalbán et al. 2013). It is thus a very precise proxy of the evolution of the stellar structure with age, even if measured in non-asymptotic conditions (Mosser et al. 2011). It allows for instance a clear distinction between a horizontal branch red giants (or red clump, RC here after) having ignited the fusion of helium in a small convective core and a red giant branch stars still having an inert helium core (Bedding et al. 2011) although both stars might have exactly the same radius and effective temperature.

The situation is still progressing in order to characterize both the core and the envelope of the red giants. Rotation of the red giant core can be measured (Beck et al. 2012, Mosser et al. 2012, Deheuvels et al. 2014). Recently, Gehan et al. (2018) could perform the automatic measurement of the rotation for about 900 red giants on the RGB. Several works are under way to make automatic and extend to a large number of stars the measurements of the pseudo-periodic perturbation of the pulsation frequency pattern characterizing the acoustic depth of the outer convective zone (Vrard et al. 2015); in parallel, further information on the convection comes from the measurement of the mode lifetimes (Vrard et al. 2018) and of the luminosity bump (Khan et al. 2018).

The potential of asteroseismology illustrated above on red giants can be extended to a large number and variety of stars. The measurement and use of Δν and ν$_{max}$, for instance, can be straightforwardly applied to other solar-type pulsators (stars showing, as the Sun and red giants, pulsations excited by the near surface convection) which cover a large part of the HR diagram. This includes all main-sequence stars cooler than early F, subgiants, and AGB stars.

**Beyond solar-type oscillations**, many classical pulsators spanning various mass ranges and evolution stages in the HR diagramme provide in their pulsation spectra patterns like $\Delta\nu$ or individual eigenfrequency values indicative of the mean star density. This is well known for Cepheids, RR Lyrae, but it is also potentially the case for main-sequence pulsators of different mass. For instance, δ Scuti stars are 1.5-2.5 stars in the main sequence and early subgiant phase. The analysis of their oscillation based on about 1800 of them observed with CoRoT has confirmed the existence of a regular pattern (Garcia Hernández et al. 2013, Paparo et al. 2016, Michel et al. 2017) and this frequency spacing, very similar to $\Delta\nu$ has been shown to be a precise indicator of the mean density (Garcia Hernández et al. 2015, 2017). The effective extension of the seismic characterization is thus to large extent a matter of collecting precise photometric time series of large enough samples of these various pulsators in order to achieve the comprehensive analysis of their pulsation spectrum.

**Rotation and activity:** Several thousands of light-curves obtained with CoRoT and *Kepler* have revealed the signature of rotational modulation attributed to starspots (Lanza et al. 2010, De Medeiros et al. 2013, Garcia et al. 2013, Bravo et al. 2014, Leão et al. 2015). In addition to providing a measurement of the rotational period free of the sin *i* projection factor, these light-curves revealed that the signature of activity is not only found in late F-G stars, for which magnetic fields and starspots were expected as in the Sun, but also in intermediate A and B type stars (Savanov 2019, David-Uraz et al. 2019, Balona 2019). This interpretation is supported by the discovery of the weak signature of magnetic field in A and B type stars (Lignières et al. 2009, Böhm et al. 2015, Blazère et al. 2016). Since the surface rotation rate of a star is in first place a function of its mass and evolutionary state (at least for low-mass stars), gyrochronology (e.g. Barnes 2007; Meiböm et al. 2011) takes advantage of these measurements of the stellar rotation rate to produce an estimate of age.

**Granulation:** The light-curves of red giants have also unveiled the presence of a signal characterizing the stellar granulation, which is a superficial signature of convection inside the stellar envelope. As for the seismic indices, it is possible to extract in rather easy way the characteristic parameters of stellar granulation. These parameters also obey scaling relations, which in addition to seismic scaling relations, provide information about the surface layers of the star. In particular, the CoRoT and *Kepler* observations have revealed the existence of a universal scaling relation between the characteristic time-scale of stellar granulation and surface gravity (Kallinger et al. 2010, Mathur et al. 2011, Bastien et al. 2013, Kallinger et al. 2016); in absence of detectable oscillations, this scaling provides an estimate of star surface gravity.

## 2b) Specific observational constraints:

A practical difficulty exists however to design a unique large sky coverage survey for seismic characterization of stars in our Galaxy. It resides in the different pulsation time scales and amplitudes between these evolution stages. The Main Sequence objects for instance are the more demanding in noise level, with amplitudes of the order of 1 ppm $\mu Hz^{-1/2}$ and in sampling cadence (~1min or less), but a couple of months are enough to resolve their eigenfrequency pattern and extract seismic indices. At the other end of the evolution sequence, AGB stars have more comfortable amplitudes (~100 ppm $\mu Hz^{-1/2}$) and sampling cadence (a few days sufficient) but the frequency resolution necessary to measure their large separation $\Delta\nu$ and $\nu_{max}$ requires long durations, of the order of 7 to 10 years (see Soszynski et al. 2004, Soszynski et al. 2009, Mosser et al. 2013, Takayama et al. 2013)

This is why it is reasonable to plan a few distinct surveys, optimized each for a given domain of the HR diagramme.

As we will show in Sect.4, PLATO represents presently the best opportunity to lead a substantial survey on Main Sequence solar-type pulsators on about half the sky. At the other extremity of the HR diagramme, LSST is very promising for a survey of AGB stars on about half the sky too.

**In the present white paper, we consider the intermediate domain between these two extremes. We focus on what we can learn about our Galaxy with a survey able to characterize seismically all stars from subgiants, to red giants and early AGB in a few kpc around our Sun.**

### 3) How our understanding of our Galaxy will increase with a fast cadence long duration variability survey of red giant stars over a few kpc

The understanding of our Galactic surrounding and of our Galaxy as a whole is experiencing an unprecedented progress under the influence of the *Gaia* mission (*Gaia* Collab. et al. 2018a) combined with ongoing and soon-to-start large spectroscopic surveys, such as APOGEE (Majewski et al. 2017), WEAVE (Dalton et al. 2012) or 4MOST (de Jong et al. 2019). In this context and in the 2035-50 perspective, we illustrate how this understanding can be dramatically improved with information about mass, age, rotation, etc…brought by a *variability survey* (Anders et al. 2017ab, Rendle et al. 2019, Valentini et al. 2019).

In order to illustrate in a practical way the interest of the information coming from variability for the study of our Galaxy, we focus here on what can be done with red giant stars and give a non-exhaustive list of breakthrough studies that could be conducted with a survey like *Chronos* (described in Section 4).

Generally speaking, **understanding our Galaxy** is relying on a precise census of the positions, kinematics, chemical composition and age of its stars. The ongoing *Gaia* programme is building such a census with measurements of unprecedented precision for over a billion of stars over a large part of our Galaxy. The mass and age estimate however is presently the weak link of the chain. Most stellar masses and ages used in Galactic astronomy are still based on fitting stellar positions in CMD diagrams with stellar evolution model isochrones (e.g., Sanders & Das, 2018; Gallart et al. 2019). The *Gaia* survey and the complementary large spectroscopic surveys will bring refined luminosity (L) and effective temperature ($T_{eff}$) and chemical composition estimates to make this comparison. However, there are fundamental limitations to this approach:

-one limitation resides in the reliability of the theoretical evolution sequences used for comparison. This is a reason why improving our understanding of stellar structure and evolution is important and one of the ways to do it right now is stellar ensemble asteroseismology. This is why time domain surveys like CoRoT, *Kepler*, K2, TESS or PLATO always represent an opportunity to improve the understanding of stellar structure and evolution and thus improve theoretical evolution sequences. This objective is included in PLATO scientific objectives and we can foresee great advances on the reliability of theoretical sequences for the Main Sequence cooler than early F and for the subsequent subgiant and red giant phase.

-the second limitation comes from the fact that, depending on the region considered in the HR diagramme, Teff and L values characterizing evolution sequences show more or less dependency on age, mass, and chemical composition. The case of red giants is very illustrative of this point. The red giant parts of the evolution sequences are clustering in a small range of Teff. Gaia brings very precise measurement of L (thus of R, for a known Teff), but the Teff measurement, as precise as it can be, cannot bring a strong constraint on the mass estimate even if the chemical composition is assumed to be perfectly known. As a result, the constraint on the age is very poor, since for these red giants, mass is the predominant key factor for age on the RGB (Miglio et al. 2017).

As commented in Sect.2, one of the first information brought by seismology is the star mean density. This is true for various types of pulsators and especially for red giants where the combination of the two seismic indices $\Delta\nu$ and $\nu_{max}$ with an estimate of $T_{eff}$ (which does not even have to be very precise, since the dependency in $T_{eff}$ is small) leads straightforwardly to the mass estimate which in turn provides a precise age estimate for several tens of thousands of red giants already.

A non-exhaustive list of breakthrough Galactic studies that could be conducted with a survey like *Chronos* in a few kpc around the Sun is the following:

— **Chemical evolution of the Milky Way.** Detailed relations between age - chemical abundance ratios of a variety of different elements (e.g., Snaith et al. 2015) have proved to be fundamental for constraining the chemical evolution of our Galaxy, but we are just at the beginning of using them, because of the difficulty to obtain ages with the useful precision. Yet, even with the precision provided by stellar isochrone age dating, these relations show remarkably tight correlations between age and chemical element abundance ratios that need to be confirmed with better ages, in particular for stars older than 8 Gyr (e.g., Haywood et al. 2013, Nissen et al., Spina et al. 2017). These relations in turn can be used to constrain the star formation history of the Milky Way, its gas accretion history, nucleosynthetic yields, but also dynamical mechanisms such as stellar radial migration triggered by a combination of the effect of external and internal perturbations (such as the bar and spiral arms, e.g., Minchev & Famaey 2010).

— **Measuring the Milky Way Star Formation History**. The measurement of the Milky Way star formation history is not reflected in the age distribution of the stars in the solar vicinity, because most old stellar populations are confined to the inner Galaxy (e.g. Bovy et al., 2012), beyond the reach of presently available stellar age dating methods. Today, our best prospect to measure the SFH of our Galaxy is to measure the stellar mass density of populations as a function of alpha abundance, and to relate these abundances to a well-calibrated age scale, determined locally.
Ongoing or upcoming large spectroscopic surveys (see above) will provide deep spectroscopic sampling of all stellar populations throughout the Milky Way, on distances of several kpc, allowing the determination of stellar densities as a function of (alpha) abundances, or mono-abundance populations (Bovy et al. 2012, Mackereth et al. 2017). Because of the relations linking age to alpha element abundances, chemical abundances can be used as proxy for stellar age on very large distances, allowing us to determine the amount of stellar mass as a function of age in the Galaxy, or the Star Formation History. Ultimately, the validity of the determination of the Star Formation History will however depend on how accurately chemical abundances can be linked to an accurate age-scale provided by the proposed survey. It will also allow us to better understand how the Milky Way disk was built (inside-out?) and crucially to compare the Milky Way history with other galaxies of the same mass.

— **Understanding the early phases of our Galaxy.** Precise ages, or high temporal resolution, are also necessary to understand the sequence of events that built the early Galaxy. Recent *Gaia* results suggest for instance that a large fraction of the oldest stars in the Galaxy have hot orbits aligned with the Galactic plane (Sestito et al. 2019), and that the stellar halo is almost fully accreted (e.g., Di Matteo et al. 2018). Moreover, it appears that these accretion events might in turn have played an important role in shaping the in-situ disk components of the Galaxy, in particular the thick disk component (e.g., Helmi et al. 2018). The formation of the first stars, the rapid early chemical enrichment, the dissipation of the gas in a disk configuration, the accretion and merger of satellite galaxies, all these event occurred within a few Gyr, and a high temporal resolution, complemented with spectroscopic kinematic information is thus needed to disentangle the different components (in situ or accreted) and reconstruct the early history of our Galaxy. This can be done with *Chronos* with red giants in a few kpc around the Sun.

— **Kinematics of the Galactic disk**. At magnitude brighter than G~ 12, *Gaia* provides the best performance in terms of stellar radial velocities and proper motions, yielding extremely detailed kinematics of the solar vicinity (d<2-3kpc for red giants). While accretion events might have played an important role in the early shaping the in-situ disk components of the Galaxy, in particular the thick disk component (see above), it appears that more recent accretions are still influencing disk dynamics today. One of the main findings of *Gaia* has been a phase-spiral when looking at the phase-space surface of section of height vs. vertical velocity for local stars (Antoja et al. 2018), which could not be revealed before due to uncertainties on parallaxes and proper motions. This clear out-of-equilibrium signature in the disk kinematics has been suggested to be the consequence of the passage of the Sagittarius dwarf through the plane of the Milky Way (Antoja et al. 2018, Laporte et al. 2019), but other explanations based on the buckling of the Galactic bar have also been proposed (Khoperskov et al. 2019), although the most likely explanation is a complex interplay between such external and internal mechanisms. What is certain is that it means that the hypothesis of equilibrium for modelling the Galactic disk can only be used as a basis for perturbation theory. A crucial observable allowing to disentangle the various proposed scenarios is the shape of this phase-spiral as a function of stellar ages (e.g., Laporte et al. 2019), which *Chronos* will provide. Moreover, the current structure of the Galactic bar itself is also constrained from *Gaia* data on disk kinematics in the Solar vicinity (e.g., Monari et al. 2019), but again, the dynamical effects of non-axisymmetric perturbations on stellar populations of various ages would allow to disentangle better the effect of the bar and spiral arms, and also to constrain the age of the Galactic bar itself, which is still largely unknown.

The *Chronos* sample would thus allow a detailed study of the kinematics of the disk as a function of age, and would contribute to the understanding of the various effects of external and internal perturbations, such as resonances of the bar and spiral arms and their signature in the solar vicinity. Coupled with available spectroscopy, it would also permit to study the effect of radial migration (e.g., Minchev & Famaey 2010), which intensity and effect on stars in our neighbourhood is still largely debated (e.g., Minchev et al., 2011, Hallé et al., 2015). More generally, a detailed characterization of the phase-space distribution of red giant stars as a function of age and chemical composition should allow to help re-constructing the global secular evolution of the Galaxy by making use of a dressed Fokker-Planck approach as presented in, e.g., Fouvry et al. (2015).

The above four examples of ground-breaking studies in the context of Galactic dynamics and archaeology are limited to red giants for which we already have developed a sufficient understanding to build precise seismic diagnostics and seismic characterization. Beyond red giants, as we explain in Sect.2, we can foresee natural extension of the application of seismic characterization on large portions of the sky to main sequence solar-type pulsators (with PLATO) and to very luminous AGB stars (with LSST). We also mentioned in Sect.2 the interest and possibility to extend this approach to main sequence classical pulsators. Because their pulsations are not so different in characteristic time scales, in amplitude and in necessary resolution frequency, these objects would also be covered by the survey considered in Sect.4.

Applying seismic characterization to pulsating stars with mass between 1.5 and 15 on the Main Sequence and subgiant branch would enrich the study of our Galactic neighbourhood (not as deep as for red giants which are intrinsically very luminous, but on a few hundreds of pc and open new perspective also because of the much shorter time scales of evolution toward high masses.

## 4) *Chronos* - An all-sky long-term seismic survey to characterize half a million red giant stars up to 1.7 kpc around the Sun

Here we show that it is possible to plan an all-sky, fast cadence, long duration photometric survey satisfying the needs of the scientific objective introduced in Sect.3. The concept presented in Sect.4a is of course only illustrative and possible ameliorations are left aside for a later phase. **As it is, we would expect this project to match the requirements for a medium size ESA mission.**

### a. A possible instrument and mission profile

**Mission concept - The instrument and the spinning platform:**

A platform is carrying eight identical elementary cameras, each covering a 11.25x11.25 deg$^2$ field of view (FOV). Cameras are grouped by two, so that at any time they are staring at four fields on the sky, 22.5x11.25 deg$^2$ each (see Fig.1).

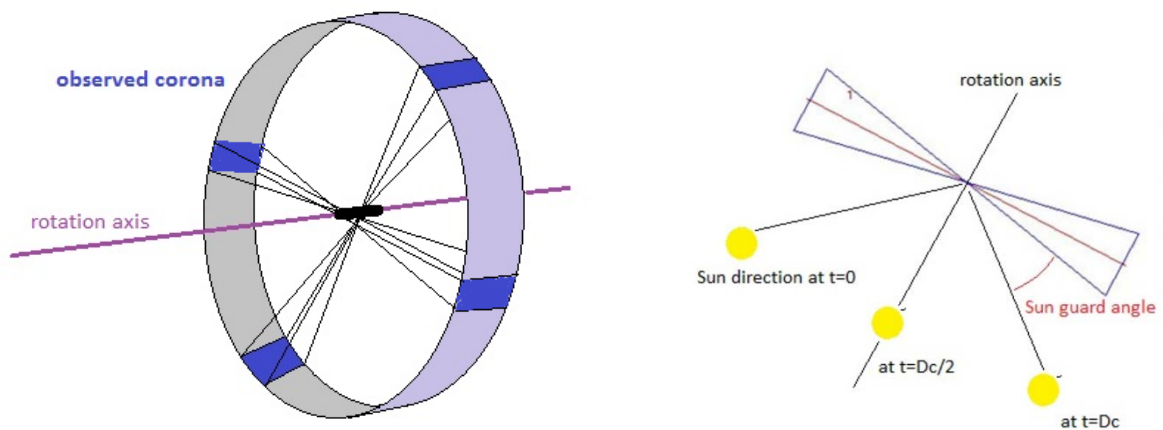

**Fig.1 left:** The 22.5x360 deg$^2$ sky corona swept by the instrument every Ts=2 minutes (rotation period of the platform 8 min). **Right:** the same corona is observed during Dc, keeping the Sun away by the guard angle GA.

The platform is spinning (Fig.1) around the same rotational axis during Dc (here we take Dc=3.75 months) and the detectors "sweep" regularly a 22.5x360 deg$^2$ corona on the sky. The rotation period $P_{rot}$ (here $P_{rot}$=8 min) ensures that each star of this corona is observed regularly with a sampling time Ts=$P_{rot}$/4 (here Ts=2 min).
The image of each star is crossing the focal plane in 15 s and thus one CCD in 7.5 s. The CCDs are read in TDI mode (time delay integration mode, as for *Gaia*, where images cross the CCD in 4.5s), the charges being drifted at the same speed as the star images.

A first optical sketch has been drawn considering, for each camera, a dioptric (6-lenses) system with a 30 cm objective, illuminating a focal plane with four CCDs (4.5k x 4.5k pixels, 18μm/px, similar to PLATO CCDs) and a point spread function within 4 pixels anywhere on the square FOV 11.25x11.25 deg$^2$.

| *Chronos* main figures | Sizing |
|---|---|
| Entrance pupil | 30 cm |
| Effective pupil* | 10.6 cm |
| Number of detection block | 4 |
| Number of cameras / block | 2 |
| Observation cadence | 2 min |
| FOV of a camera | 11.25° |
| Size of each corona field | 22.5° x 360° |
| Observation shift between consecutive corona | 112.5° |
| Observation length per corona | 3.75 months (2 times in nominal lifetime) |
| Mission nominal lifetime | 5 years |

**Tab.1** Summary of the main figures of the mission concept in its present version. *The effective pupil takes into account that each star of a given corona is observed by a camera regularly but during 1/8 of the time.

**Mission concept - orbit and observational strategy:**

We considered a platform located at Sun-Earth L2, observing the same corona on the sky during Dc=3.75 months and change from one corona to the next. With a rotation axis chosen to point toward the Sun at Dc/2, the Sun is kept away from the FOV by a guard angle GA (here GA=22.5 deg).

**Phase 1 (2.5 years):** by shifting every Dc (3.75 months) the rotation axis by 112.5 deg (112.5=3.75x30 being the angle described by the Sun in 3.75 months), it is possible to cover the whole sky with 8 successive coronas within 30 months. For comparison, covering the whole sky by shifting every 3 months a pointed instrument with a FOV of 1000 deg$^2$ requires more than 10 years.
We notice that beside this, the two regions of the sky where the 8 coronas overlap will be observed a longer time (up to the total length of the mission for the continuous viewing zones CVZ, two symmetric disks of diameter 22.5 deg centred at the ecliptic poles).

**Phase 2 (+30 months → 5 years, nominal duration of the mission):** with 16 runs of duration Dc, the whole sky will be observed twice and each target observed two times Dc, separated by 7 Dc (here 2x3.75 months separated by ~2 years and 2 months).

**Phase 3 (possible extension: +30 months → 7 years and 6 months):** with 24 runs of duration Dc, the whole sky will be observed 3 times Dc, separated by 7 Dc (here 3x3.75 months separated by ~2.years and 2 months each)

**Performances in terms of seismic indices and seismic diagnostics performances:**

These estimates result from tests and scalings based on CoRoT and *Kepler* data. These tests show that seismic indices of red giants could be measured on discontinuous periods 2xDC or 3xDC separated by ~2 years as considered here with the following performances (cf Tab.2).

**In phase 2 (5 years, nominal mission):** we measure seismic indices Δν and ν$_{max}$ for all red giants (RGB, RC, early AGB) and up to 1.7 kpc (m$V$ ≤11) and for part of the subgiants up to ~200pc (m$V$ ≤10), i.e. approximately 500 000 stars spread over the whole sky. We also measure indices ΔP for all the red giants around the RC, where this index is of crucial interest to disentangle stars on the RGB from stars on the RC.

**In phase 3 (7.5 years, extended mission):** Δν and ν$_{max}$ and ΔP are measure with an increased precision for all red giants and for subgiants with 8 ≤ m$V$ ≤ 11. .

**The Continuous viewing zone:** The survey of *Chronos* gives access to a continuous viewing zone (CVZ), two cones of angular diameter 22.5 deg, about the size of the of TESS CVZ (24 deg) and in the same region as TESS (Eccliptic pôles). The scientific yield for stars observed 5 years in the CVZ will outpass all projects, with many more objects to be observed much longer in all regions of the HR diagram. For these stars, the most precise seismic analysis will be performed: analysis of the signatures of helium, rotation and convection for main-sequence stars; core properties and rotation profiles for evolved stars.

| Observation duration | | MS type G | MS type F | Subgiant | Early RGB | Red clump | Early AGB |
|---|---|---|---|---|---|---|---|
| 3.5 months | 1x | 8.0 | 8.8 | 9.5 | 10.5 | 12.0 | 11.1 |
| | 2x | 8.6 | 9.5 | 10.2 | 11.0 | 13.3 | 12.4 |
| | 3x | 9.2 | 10.2 | 10.8 | 11.5 | 14.6 | 13.7 |
| 5 years | CVZ | 11.3 | 12.4 | 12.9 | 13.5 | 16.0 | 15.1 |

**Tab.2** limit m$_V$ magnitude for Δν and ν$_{max}$ measurement, for different observation duration and different regions of the HR diagram (Mosser et al. 2019). Green indicates the matching with the 8-11 m$_V$ range observable with *Chronos* in its present version.

### b. Programmatic context: Comparison with existing and planned space and ground based projects.

As mentioned previously, the need for photometric surveys covering a large part of the sky over long duration has been taken into account as much as possible in various existing and planned space and ground-based projects. However, there are strong practical difficulties to optimize observation for such a large range of contradictory constraints. We show here that none of these projects is suited for the objectives fixed in our present scientific proposal. This is understandable since these projects usually had to optimize in first place other scientific objectives with other instrumental constraints, like the search for exoplanets.

**CoRoT, *Kepler* and K2**: CoRoT and *Kepler* have brought wonderful data for red giant stars, because the duration of the runs was at least several months and up to years in the case of *Kepler*. But this long duration was paid in both cases by a limited coverage of the sky. The K2 extension of the *Kepler* mission allowed to enlarge the sky coverage considerably, (although it was still not complete) but the duration of the runs was limited to 80 days, which is not enough to derive the seismic indices needed in the present project.

**TESS:** With 86% of the all-sky accessible, TESS (Ricker et al. 2015) is the project offering presently the largest sky coverage. However, the primary goal of TESS is to discover planets smaller than Neptune that transit stars bright enough to enable follow-up spectroscopic observations that can provide planet masses and atmospheric compositions, with short revolution periods and most of the sky (63%) will be observed only 27 days within the 2-year nominal mission and only 23% will be observed 54 days or more. This duration is too short to measure seismic indices for red giant stars.

**PLATO:** PLATO mission profile (Rauer et al. 2016) is organized around 2 long duration fields totaling 4400 deg$^2$ to be observed during 2 to 3 years each. Then, it is planned to complete the observations with a few step and stare fields of shorter duration in order to cover about half the sky. Considering the scientific objective considered here, the two long duration field will bring valuable data for red giants, but on a limited fraction of the sky. The step and stare fields, if they have to increase this sky coverage significantly, say to half the sky, will hardly reach more than 2.5 months each. This would be a great survey in terms of solar-type Main Sequence pulsators, but this duration is too short to properly estimate seismic indices like $\Delta P$ for red giants.

**LSST**: The LSST survey (LSST Science Collaboration et al. 2009) will cover about half the sky (southern hemisphere) with two consecutive 15s measurements every 3 days at best. Such a sampling rate is not suited for the red giant stars considered in the present project. On the other hand, it will be very efficient for AGB stars or very large red giants near the tip of the red giant branch, which require very long duration observations (7-10 years) as planned for LSST.

***Chronos*:** The present project, offering an all-sky survey with long duration (2x3.75 months within the 5-year nominal mission) will bridge the gap left between LSST and PLATO, bringing seismic indices for solar-type pulsators from subgiants to early AGB stages (~half a million red giants and subgiants), plus classical pulsators on the Main Sequence ($\delta$ Scuti, $\gamma$ Doradus, $\beta$ Cephei and SPB stars).

| Project | Sky coverage (%) | Observation length and sky fraction | | Cadence of observation and number of observed stars | |
|---|---|---|---|---|---|
| | | Min | Max | SC | LC |
| CoRoT | 0.20 | 1 month | 5 months | 32 sec ~ 1k | 8.5 min ~ 150k |
| *Kepler* | 0.25 | 1 month | 4 years | 2 min ~ 1k | 30 min ~ 150k |
| K2 | 2.5 | 80 days | | ~ 5k | ~ 2M |
| TESS | 86 | 27days (63%) 54days (23%) | 1 year (CVZ ~ 2 %) | 2 min ~ 200k | 30 min ~1M |
| LSST | ~ 50 | 10 years | | ~ 3 days | |
| PLATO | ~ 50 | 2 months (50%) | 3 years (10%) | 1 min | 10 min ~1M |
| *Chronos* | 100 | 2 x 3.75 months (100%) | > 5 years (CVZ ~ 2 %) | 2 min 1M | |

**Tab.3** Comparison of *Chronos* main figures with other past and expected projects

### 5) Possible instrumental improvements and technology challenges

The technology considered here to support the *Chronos* Project is inherited from the various photometric surveys realized or designed over the last decades (CoRoT, *Kepler*, TESS, PLATO) and from the *Gaia* mission for the spinning platform and the time delay integration (TDI) readout of the CCDs. We thus are reasonably confident in the possibility to reach the performances anticipated, even if there is probably room to optimize the instrumental concept considered here.

**An orbit allowing longer individual continuous runs (beyond 3.75 months):**

One interesting possible significant improvement would be to find an orbit allowing uninterrupted observations longer than the 3.75 months considered in the present version. Compared with the present 2 x 3.75 months sequences, an uninterrupted sequence would allow not only to measure ∆P but also ∆π1, the asymptotic period spacing which is a very sensitive indicator for evolution but requires the measurement of g-dominated mixed modes with long lifetime. These modes appear as coherent pulsations and are unresolved in the Fourier spectrum even for long durations of several years. In such a case, continuous runs of e.g. 7.5 months would be better than discontinuous (2 x 3.75 months).

Increase the duration would require a change in the present concept that is challenging. Among possibilities we could think of:

- an orbit eclipsing the Sun: the length of individual runs DC* (Dc*=3.75 months in the present setup) is determined by the will not to have the Sun coming closer than GA=22.5 deg from the observed corona. This constraint could be relaxed (and Dc* increased without limitation) if it

- were possible to operate an orbit where the platform is in the shadow of the earth (or of an artificial shield) at least during the 1-2 months when the Sun crosses the corona (possibly all the time).
- positioning the platform further from the Sun. For instance, at Mars L2 point (or any where at Mars distance from the Sun): this would naturaly increase Dc* by a factor ~2 (year_Mars/Year_Earth=1.9) without changing the mission observational strategy.

**An adaptation of the TDI mode allowing an increase of the angular size of the coronas:**

In the present design, corona are limited to 22.5 deg width. Enlarging this width (on the side opposite to the Sun) would improve the sky coverage performances. This increase however would lead stars images to describe curved trajectories on the CCDs instead of staying in the same ~4 columns. The solution consisting in increasing the number of columns considered for each star is a possibility but rapidly becomes unsatisfying (overlap of the images of different stars).

A possibility would be:

- adapt the TDI readout mode to take into account the curvature of the trajectory of star images when crossing the CCD. This would mean attributing to each star the photoelectrons collected in different columns as time varies. This approach has to be tested, but the fact that we stay on one given corona for several months should help since the trajectories are not expected to change at first order. We note that this option would also relax the constraints on the alignement of the CCDs (1' in the present design to keep the trajectory within the same 5 columns) and on the stability of the orientation of the rotation axis (also 1' in the present design).

## 6) Summary and conclusion in a broad scientific perspective

The period 2035-50 considered for the Voyage ESA long-term plan lies at the crossroad of several major progresses in the characterization of the stellar content of our Galaxy. *Gaia* and the various spectroscopic large surveys are building up an unprecedented census in terms of astrometric, kinematic and chemical properties of the Galactic stellar populations. Within a decade, precise measurements of these properties will thus be available for hundreds of millions of stars over a large part of our Galaxy. Meanwhile, stellar time domain surveys initiated by CoRoT and *Kepler* and continued by space missions like TESS and PLATO or ground based projects like LSST will have brought stellar seismology and stellar seismic characterization to a high level of maturity. The synergy between seismic information and the astrometric one and the great interest for Galactic population studies has already been illustrated with tens of thousands of red giant stars and we can foresee its extension to other types of stars. This is why we are convinced that an all-sky, high-cadence, long-duration variability survey will be a top scientific priority in the 2035-50 period.

The *Chronos* concept presented here, constitutes a time-domain extension of *Gaia*. It will allow mass and age estimates for half a million red giants within 1.7 kpc from the Sun and shed a new light on our understanding of the Galactic dynamics and archaeology.

In terms of characteristic time scales, *Chronos* will bridge up the gap between PLATO and LSST. It will outpass all previous time-domain projects in terms of sky coverage (100%) and duration (2 x 3.75 months within the 5-year nominal mission).